\documentclass[12pt, reqno]{amsart}
\usepackage{amsmath, amsthm, amscd, amsfonts, amssymb, graphicx, color}
\usepackage[bookmarksnumbered, colorlinks, plainpages]{hyperref}
\hypersetup{colorlinks=true,linkcolor=red, anchorcolor=green,
citecolor=cyan, urlcolor=red, filecolor=magenta, pdftoolbar=true}

\newtheorem{theorem}{Theorem}[section]

\theoremstyle{definition}

\theoremstyle{remark}
\newtheorem{remark}[theorem]{Remark}
\numberwithin{equation}{section}
\input{tcilatex}

\newcommand{\MPAG}{Math. Phy. Ana. Geom.}

\begin{document}

\setcounter{page}{1}

\title{Numerical results a quantum waveguide with Mixed boundary conditions}

\author{M. Raissi$^1$}
\address{$^{1}$ D\'epartement de Math\'ematiques, Facult\'{e} des Sciences de Moanstir. Avenue de
l'environnement 5019 Monastir -TUNISIE.\newline Laboratoire de
recherche: Alg\`{e}bre G\'{e}om\'{e}trie et Th\'{e}orie Spectrale:
LR11ES53}
\email{\textcolor[rgb]{0.00,0.00,0.84}{raissi.monia@yahoo.com}}
\begin{abstract}
This article is devoted to the numerical study of the existence of
the eigenvalues of the Hamiltonian describing a quantum particle
living on three dimensional straight strip of width $d$ in the
presence of an electric field of constant intensity $F$ in the
direction perpendicular to the electron plane. We impose Neumann
boundary conditions on a disc window of radius $a$ and Dirichlet
boundary conditions on the remaining part of the boundary of the
strip.
\end{abstract} \maketitle
\section{Introduction and the model}
%%%%%%%%%%%%%%%%%%%%%%%%%%%%%%%%%%%%%%%%%%%%%%%%%%%%%%%%%%%%%%%%%%%%%%%%%%%%%%%
The system we are going to study is given in Figure 1. This system
is based on the work of Najar et al in \cite{reff} where it is
proved that such system admits a discrete spectrum below its
essential spectrum we are interested on numerical results of some of
untreated cases in the mentioned reference. Here we our computation
are based on Mathlab and Maple.\\  We consider a quantum particle,
this leads to the study of an Hamiltonian which we denote by
$H_a(F)$, whose motion is confined to a pair of parallel plans of
width $d$. For simplicity, we assume that they are placed at $z=0$
and $z=d$. We shall denote this configuration space by $\Omega$
\[
\Omega=\Bbb{R}^2\times [0,d].
\] We suppose that the particle is a
fermion of a nonzero charge $q$. We also assume that it is under
influence of a homogeneous electric field of an intensity $E$, we
denote $\displaystyle F:=Eq$. Without loss of generality we shall
suppose in the following that $\displaystyle F\geq0$ and that the electric field is perpendicular to the electron plane.\\
Let $\displaystyle\gamma(a)$ be a disc of radius $a$, without loss
of generality we assume that the center of $\displaystyle\gamma(a)$
is the point $(0,0,0)$;
\begin{equation}
\gamma(a)=\big\{(x,y,0)\in \Bbb{R}^3;\ x^2+y^2\leq a^2\big\}.
\end{equation}
We set $\displaystyle \Gamma=\partial\Omega\diagdown \gamma(a)$. We
consider Dirichlet boundary condition on $\Gamma$ and Neumann
boundary condition on $\displaystyle\gamma(a)$.

\begin{figure}[ht]
\centering
\includegraphics[width=1\columnwidth]{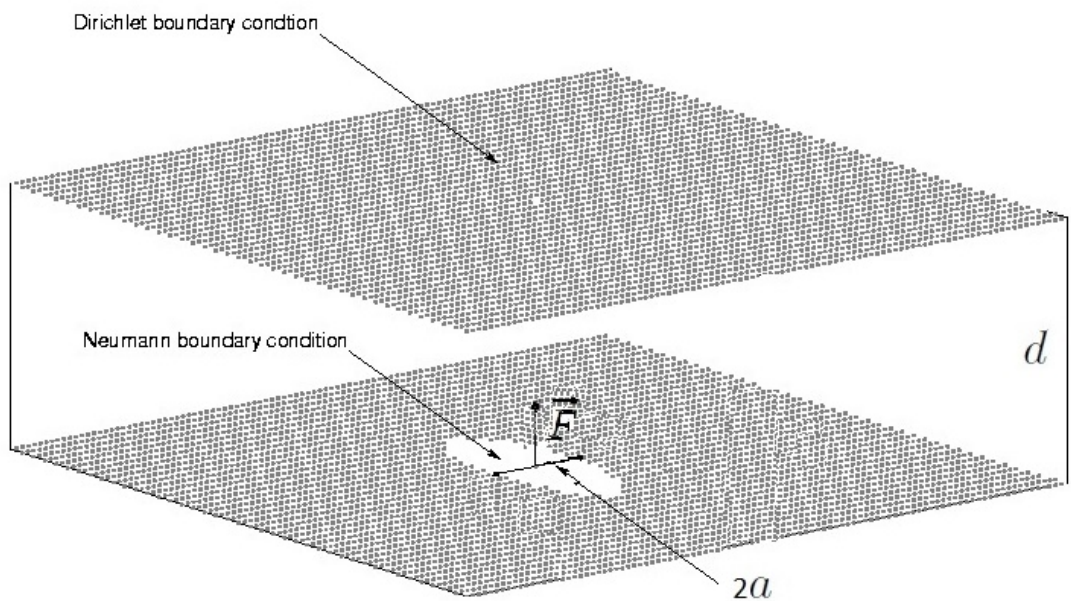}
\caption{\label{fig1} The waveguide with a disc window and two
different boundary conditions with orthogonal electric field.}
\end{figure}
\bigskip
%%%%%%%%%%%%%%%%%%%%%%%%%%%%%%%%%%%%%%%%%%%%%%%%%%%%%%%%%%%%%%%%%%%%%%%%%%%%%%
\subsection{The Hamiltonian}
%%%%%%%%%%%%%%%%%%%%%%%%%%%%%%%%%%%%%%%%%%%%%%%%%%%%%%%%%%%%%%%%%%%%%%%%%%%%
 Let us define the self-adjoint operator on
$\displaystyle\mathrm{L}^2(\Omega)$ corresponding to the particle
Hamiltonian $\displaystyle H_a(F)$. This will be done by the mean of
quadratic forms. Precisely, let $\displaystyle q_a$ be the quadratic
form
\begin{eqnarray*}
  q_a[u,v] &=& \int_{\Omega}\nabla u \overline{\nabla v}+Fzu\overline{v}dxdydz
  \quad u,v\in D(q_a),
\end{eqnarray*}
where $\displaystyle D(q_a):=\big\{ u\in
\mathrm{H}^1(\Omega),u\lceil\Gamma=0\big\}$ and
$\mathrm{H}^1(\Omega)$ is the standard Sobolev space and
$u\lceil\Gamma$ is the trace of the function $u$ on $\Gamma$. It
follows that $\displaystyle q_a$ is a densely defined, symmetric,
positive and closed quadratic form \cite{Resi}. We denote the unique
self-adjoint operator associated to $\displaystyle q_a$ by
$\displaystyle H_a(F)$ and its domain by $\displaystyle D$. It is
the hamiltonian describing our system. From \cite{Resi} (page 276),
we infer that the domain $\displaystyle D$ of $\displaystyle H_a(F)$
is
\begin{eqnarray*}
% \nonumber to remove numbering (before each equation
   D&=& \big\{ u\in\mathrm{H}^1(\Omega);\quad -\Delta u\in\mathrm{L}^2(\Omega),u\lceil\Gamma=0,\frac{\partial u}{\partial z}\lceil\gamma(a)=0\big\},
\end{eqnarray*}
and
\begin{equation}\label{13}
H_a(F)u=(-\Delta+ Fz)u, \quad \forall u\in D.
\end{equation}
%%%%%%%%%%%%%%%%%%%%%%%%%%%%%%%%%%%%%%%%%%%%%%%%%%%%%%%%%%%%%%%%%%%%%%%%%%%%%%%%ùùùù
\section{Numerical computations}
%%%%%%%%%%%%%%%%%%%%%%%%%%%%%%%%%%%%%%%%%%%%%%%%%%%%%%%%%%%%%%%%%%%%%%%%%%%%%%%%%%%%%%%ù
This section is devoted to some numerical computations. Let us start
this section by giving some notations that we will use in the rest
of this work: $\displaystyle\lambda_k(H^{-,N}_a(F))$,
$\displaystyle\lambda_k(H^{-,D}_a(F))$ and
$\displaystyle\lambda_k(H_a(F))$, the $k-$th eigenvalue of
$\displaystyle H^{-,N}_a(F)$, $\displaystyle H^{-,D}_a(F)$ and
$H_a(F)$, respectively. Then, the min-max principle yields the
following
\begin{equation}\label{eqv}
\lambda_k(H^{-,N}_a(F))
\leq\lambda_k(H_a(F))\leq\lambda_k(H^{-,D}_a(F))
\end{equation}
and for $2\geq k$
\begin{equation}\label{eqv2}
\lambda_{k-1}(H^{-,D}_a(F))
\leq\lambda_k(H_a(F))\leq\lambda_k(H^{-,D}_a(F)).
\end{equation}
Thus, if $\displaystyle H^{-,D}_a(F)$ exhibits a discrete spectrum
below $\displaystyle\lambda_{0}^1$, then $\displaystyle
H_a(F)$ do as well. We mention that its a sufficient condition. \\
Let us consider the eigenvalue equation is given by
\begin{equation}\label{eq}
  H^{-,D}_a(F)f(r,\theta,z) = \lambda f(r,\theta,z).
\end{equation}
This equation is solved by separating variables and considering\\
$\displaystyle f(r,\theta,z)=R(r)P(\theta)Z(z)$.\\ We divide the
equation \eqref{eq} by $f$, we obtain
\begin{equation}\label{eeq}
 \frac{1}{R}(R''+\frac{1}{r}R')+\frac{1}{r^2}\frac{P''}{P}+\frac{Z''}{Z}-Fz = -\lambda.
\end{equation}
Plugging the last expression in equation \eqref{eeq} and first
separate the term $ \displaystyle \frac{P''}{P}$ which has all the
$\theta$ dependance. Using the fact that the problem has an axial
symmetry and the solution has to be $2\pi$ periodic and single value
in $\theta$, we obtain $\displaystyle \frac{P''}{P}$ should be a
constant $\displaystyle -m^2$ for
$m\in\mathbb{Z}$.\\
Second, we separate $Z$ by putting all the $z$ dependence in one
term so that $\displaystyle\frac{Z''}{Z}-Fz$ can only be constant.
The constant is taken as $\displaystyle \lambda_{\infty}^n$ for
$n\in\mathbb{N}$.\\
Finally, we write the equation \eqref{eeq} as a function of $R$
\begin{equation}\label{eq1}
R''(r)+\frac{1}{r}R'(r)+[\lambda-\lambda_{\infty}^n-\frac{m^2}{r^2}]R(r)=0.
\end{equation}
We notice that the equation \eqref{eq1}, is the Bessel equation and
its solutions could be expressed in terms of Bessel functions. More
explicit solutions could be given by considering boundary
conditions.\\
 The solution of the equation \eqref{eq1} is given by $
\displaystyle R(r)=cJ_{m}(\eta r)$, where $\displaystyle c\in
\mathbb{R}^{\star}$, $\displaystyle
\eta^2=\lambda-\lambda_{\infty}^n$ and $\displaystyle J_{m}$ is the
Bessel function of first kind of order $m$. \\
We assume that
\begin{eqnarray}\label{nom}
  \nonumber R(a)=0 &\Leftrightarrow& J_{m}(\eta a)=0 \\
   &\Leftrightarrow& a\eta=x_{m,k}.
\end{eqnarray}
 Where $\displaystyle x_{m,k}$ is the $k-$th positive zero of the Bessel function
 $\displaystyle J_{m}$ (see \cite{n}).\\
Then $\displaystyle H^{-,D}_a(F)$ has a sequence of eigenvalues
\cite{n,wat}, given by
\begin{eqnarray*}
     \lambda_{n,m,k} &=&  \left(\frac{x_{m,k}}{a}\right)^2+\lambda_{\infty}^n.
   \end{eqnarray*}
%By the min-max principle and \eqref{eqv}, we know that if
%$\displaystyle H^{-,D}_a$ exhibits a
%discrete spectrum below $\displaystyle\lambda_{0}^1$, then $\displaystyle H_a(F)$ do as well.\\
 % So, if the following condition
 the condition
 \begin{equation}\label{cond}
\lambda_{n,m,k}<\lambda_{0}^1,
 \end{equation}
 %is satisfied, then $\displaystyle H_a(F)$ have a discrete spectrum.\\
yields that $n=1$, so we get
\begin{equation}\label{cond11}
\lambda_{1,m,k}=\left(\frac{x_{m,k}}{a}\right)^2+\lambda_{\infty}^1.
 \end{equation}
This yields that the condition \eqref{cond} to be fulfilled, will
depends on the value of
$\displaystyle\left(\frac{x_{m,k}}{a}\right)^2$. We recall that
$\displaystyle x_{m,k}$ are the positive zeros of the Bessel
function $\displaystyle J_m$. So, for any $\lambda_a$ eigenvalue of
$\displaystyle H_a(F)$, there exists $\displaystyle
m,k,m',k'\in\mathbb{N}$, such that
\begin{equation}\label{eqafort}
   \left(\frac{x_{m',k'}}{a}\right)^2+\lambda_{\infty}^1\leq\lambda_a\leq\left(\frac{x_{m,k}}{a}\right)^2+\lambda_{\infty}^1.
\end{equation}
Using the boundary conditions, we obtain that the operators $h_0(F)$
and $h_{\infty}(F)$ have a sequence of eigenvalues
\begin{itemize}
    \item in the case of weak electric field respectively given by:
    \begin{eqnarray*}
  \lambda_{0}^n&=&\left(\frac{n\pi+\sqrt{n^2\pi^2+d^3F}}{2d}\right)^2+o(F);\quad n\in\mathbb{N}^*. \\
  \lambda_{\infty}^{n+1} &=&\left(\frac{(2n+1)\frac{\pi}{2}+\sqrt{(2n+1)^2(\frac{\pi}{2})^2+d^3F}}{2d}\right)^2+o(F);\quad n\in\mathbb{N}.
\end{eqnarray*}
    \item in the case of strong electric field respectively given by:
    \begin{eqnarray*}
     \lambda^n_0&=&-\alpha_nF^{\frac{2}{3}},\quad n\in\mathbb{N}^*. \\
      \lambda^n_{\infty}&=&-\alpha'_nF^{\frac{2}{3}},\quad n\in\mathbb{N}^*.
     \end{eqnarray*}
\end{itemize}
Where $\displaystyle\alpha_n$ and $\displaystyle\alpha'_n$ are the
$n-$th negative zeros of the Airy functions $Ai$ and $Ai'$
respectively. Consequently, we have
\begin{itemize}
    \item in the case of weak electric field respectively given by:
    \begin{eqnarray*}
  \lambda_{0}^1&=&\left(\frac{\pi+\sqrt{\pi^2+d^3F}}{2d}\right)^2+o(F). \\
  %\lambda_{\infty}^{1} &=&\left(\frac{\frac{\pi}{2}+\sqrt{(\frac{\pi}{2})^2+d^3F}}{2d}\right)^2+o(F).
\end{eqnarray*}
    \item in the case of strong electric field respectively given by:
    \begin{eqnarray*}
    \lambda^1_0&=&-\alpha_1F^{\frac{2}{3}}\simeq 2.3381F^{\frac{2}{3}}. \\
      %\lambda^1_{\infty}&=&-\alpha'_1F^{\frac{2}{3}}\simeq 1.0187F^{\frac{2}{3}}.
     \end{eqnarray*}
\end{itemize}
\begin{remark}
Using the inequality \eqref{eqafort}, for $a$ big enough, if
$\lambda_a$ is an eigenvalue of the operator $\displaystyle H_a(F)$
less then $\lambda_0^1$ then we have
$$\lambda_a=\lambda_{\infty}^1+o(\frac{1}{a^2}).$$
\end{remark}
 In the following of this section, we represent the
area of existence of the first three eigenvalues of $H_a(F)$
$\lambda_a^1$, $\lambda_a^2$ and $\lambda_a^3$ and the threshold of
appearance of eigenvalues, for the electric field of constant weak
intensity $F$
in Figure 2, and  for $F$ strong enough in Figure 3.\\
 We observe that the area of existence
of the eigenvalues of $H_a(F)$ is proportional to the intensity $F$.

\begin{figure}
\centering
\includegraphics[width=1\textwidth]{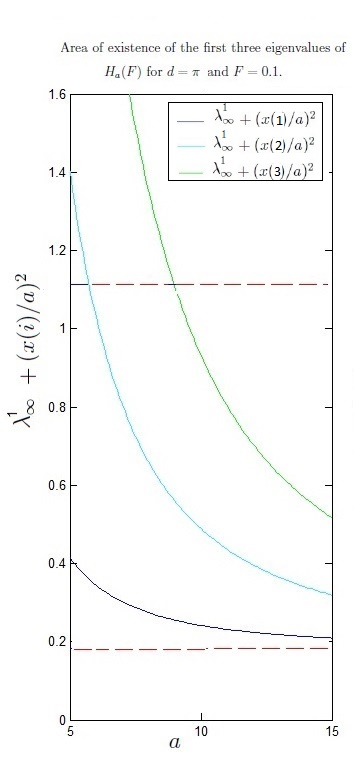}
\caption{We represent
$\left(\frac{x(i)}{a}\right)^2+\lambda_{\infty}^1$ where $x(1),
x(2), x(3)$ are the first three zeros of the bessel functions
increasingly ordered.}
\end{figure}

\begin{figure}
\centering
\includegraphics[width=1\textwidth]{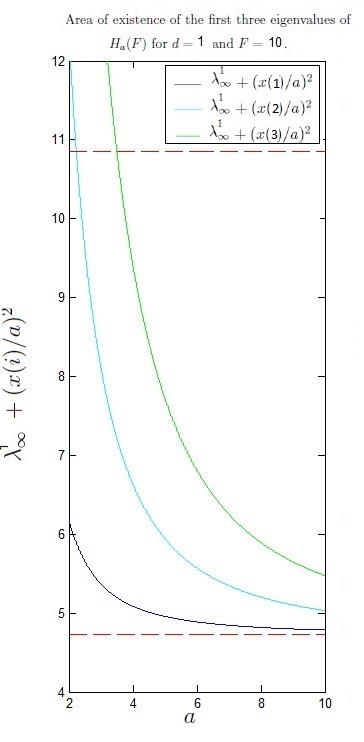}
\caption{We represent
$\left(\frac{x(i)}{a}\right)^2+\lambda_{\infty}^1$ where $x(1),
x(2), x(3)$ are the first three zeros of the bessel functions
increasingly ordered.}
\end{figure}

\newpage
\qquad In the Figure \ref{fig25}, we set the $F$ intensity of the
electric field by a low value $0.1$. We represent the curve of the
number of eigenvalues of the operator $\displaystyle H_a^D(F)$ a
function of the quotient of the radius value $\displaystyle a$ by
the width of the strip $\displaystyle d.$

\begin{figure}[h] \centering
\includegraphics[width=1\columnwidth]{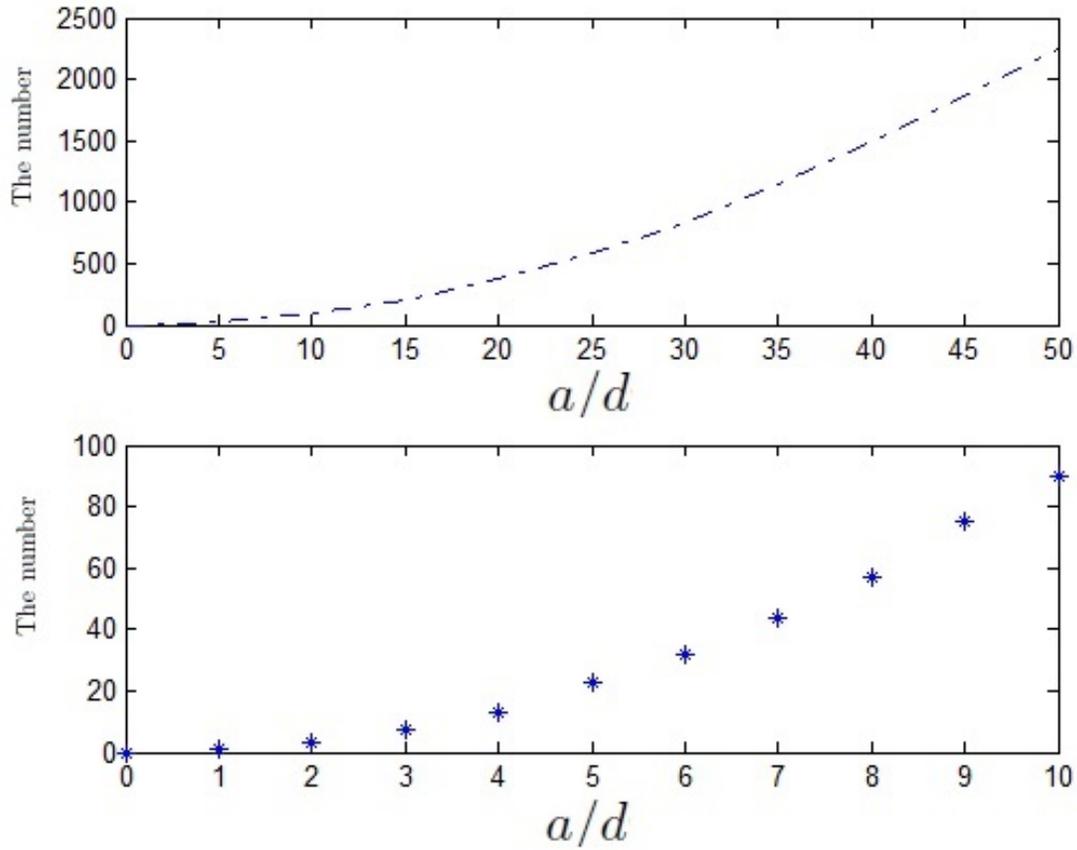}
\label{fig25}
 \caption{\label{fig25} The number
of eigenvalues of the operator $\displaystyle H_a^D(F)$ a function
of $a/d$.}
\end{figure}
\newpage
\qquad In the Figure \ref{fig26}, Similarly we set the intensity $F$
of the electric field l'intensité  du champ électriqueby a great
value $10$. We represent the curve of the number of eigenvalues of
the operator $\displaystyle H_a^D(F)$ a function of the quotient of
the radius value $\displaystyle a$ by the width of the strip
$\displaystyle d.$

\begin{figure}[h] \centering
\includegraphics[width=1\columnwidth]{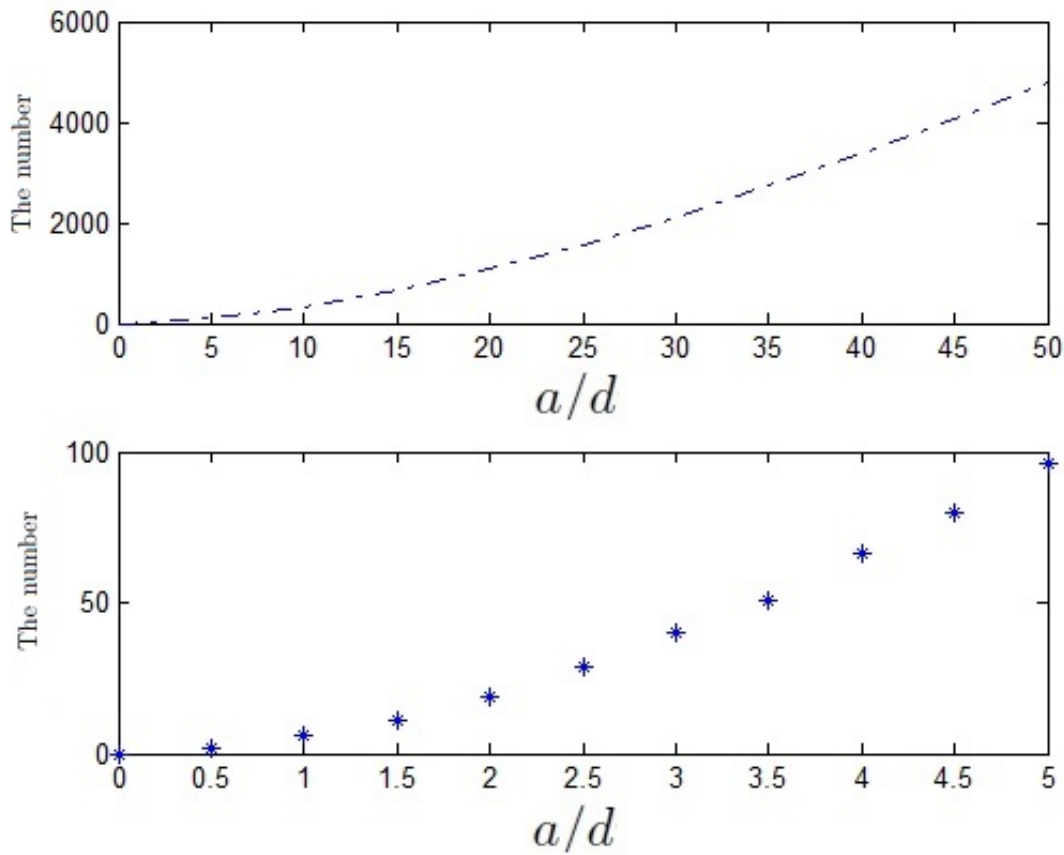}\label{fig26}
 \caption{\label{fig26} The number
of eigenvalues of the operator $\displaystyle H_a^D(F)$ a function
of $a/d$.}
\end{figure}
\newpage
\qquad In Figures \ref{fig27} and \ref{fig28}, we set the quotient
of the radius value $\displaystyle a$ by the width of the strip
$\displaystyle d$ by real $10$. We represent the curve of the number
of eigenvalues of the operator $\displaystyle H_a^D(F)$ a function
of the intensity $F$ of the electric field.

\begin{figure}[h] \centering
\includegraphics[width=1\columnwidth]{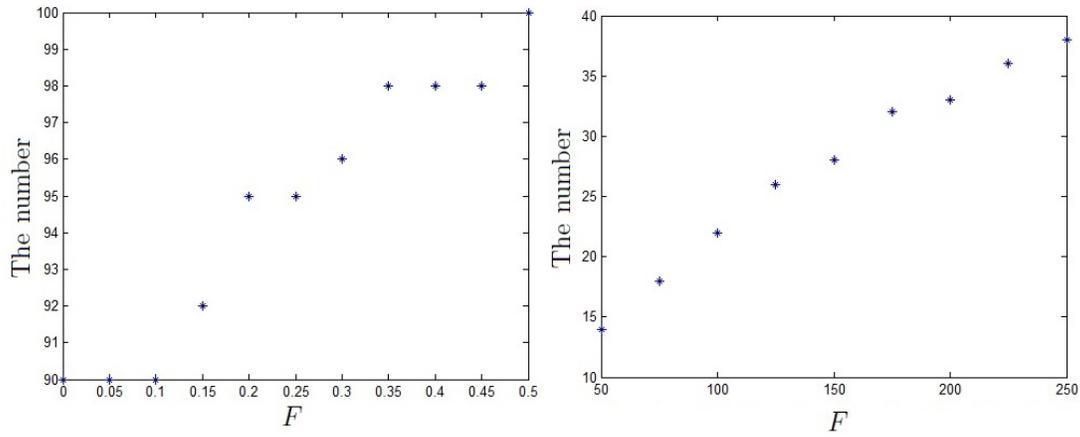}
\label{fig27}
 \caption{\label{fig27} The number
of eigenvalues of the operator $\displaystyle H_a^D(F)$ a function
of the intensity $F$.}
\end{figure}

\newpage
\begin{figure}[h] \centering
\includegraphics[width=1\columnwidth]{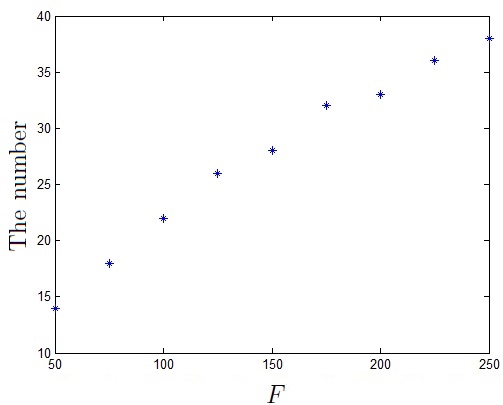}
\label{fig28}
 \caption{\label{fig28} The number
of eigenvalues of the operator $\displaystyle H_a^D(F)$ a function
of the intensity $F$.}
\end{figure}

\end{document}